\documentstyle[prl,aps,multicol]{revtex}
\renewcommand{\narrowtext}{\begin{multicols}{2}
\global\columnwidth20.5pc}
\renewcommand{\widetext}{\end{multicols} \global\columnwidth42.5pc}
\begin{document}
\preprint{}
\draft
\title{Dynamical Correlations among Vicious Random Walkers}

\author{Taro Nagao${}^{1,*}$, Makoto Katori${}^{2,\dagger}$, 
and Hideki Tanemura$^{3,\ddagger}$}

\address{
${}^{1}$Department of Physics, Graduate School of Science,
Osaka University, Toyonaka, Osaka 560-0043, Japan
}
\address{
${}^{2}$University of Oxford, Department of Physics-Theoretical Physics, 
1 Keble Road, Oxford OX1 3NP, United Kingdom
}
\address{
${}^{3}$Department of Mathematics and Informatics, Faculty of Science, 
\\ Chiba University, 1-33 Yayoi-cho, Inage-ku, Chiba 263-8522, Japan
}

\date{\today}
\maketitle
\begin{abstract}

Nonintersecting motion of Brownian particles in one 
dimension is studied. The system is constructed as the 
diffusion scaling limit of Fisher's vicious random walk. 
$N$ particles start from the origin at time $t=0$ and then 
undergo mutually avoiding Brownian motion until a finite time 
$t=T$. In the short time limit $t \ll T$, the particle 
distribution is asymptotically described by Gaussian Unitary Ensemble 
(GUE) of random matrices. At the end time $t = T$, it is 
identical to that of Gaussian Orthogonal Ensemble (GOE). 
The Brownian motion is generally described by the dynamical 
correlations among particles at many times $t_1,t_2,\cdots,
t_M$ between $t=0$ and $t=T$. We show that the most general 
dynamical correlations among arbitrary number of particles at 
arbitrary number of times are written in the forms of quaternion 
determinants. Asymptotic forms of the correlations in 
the limit $N \rightarrow \infty$ are evaluated and a 
discontinuous transition of the universality 
class from GUE to GOE is observed. 
\end{abstract}
\pacs{PACS number(s): 05.40.-a, 05.50.+q, 02.50.Ey}
\narrowtext

The vicious walk model in which many 
walkers randomly move without intersecting 
with others was introduced 
by Fisher and applied to wetting and melting 
phenomena\cite{FISHER}. Recently it has attained 
a renewed interest, since intimate relations to other 
research fields, such as the theory of Young tableaux 
in combinatorics\cite{GUT,BDJ,KRA}, asymmetric exclusion process 
(ASEP) in nonequilibrium statistical mechanics\cite{KJ}, 
polynuclear growth (PNG) model of surface physics\cite{PS1,PS2} 
and the theory of random matrices\cite{BAIK,NFV}, have been 
revealed one after another and brought progress in the study 
of these topics.
\par
One of the reigning concepts of these new applications, 
the universality class, comes from the theory of random 
matrices. Gaussian Orthogonal Ensemble 
(GOE) and Gaussian Unitary Ensemble (GUE) universality 
classes originate in the real symmetric and complex 
hermitian structures of random matrices. They also appear 
in the new applications in quite natural ways, despite 
the absence of underlying matrix structure. In the 
theory of Young tableaux, number permutation 
and involution correspond to GUE and 
GOE, respectively\cite{BR1}. In ASEP, the initial and boundary 
conditions demarcate the universality classes of the 
current fluctuation\cite{BR2,PSASEP}. The height fluctuation of 
surface growth on a flat substrate belongs to the GOE class 
and a droplet growth is described by GUE\cite{PS2}. 
\par
How the universality classes appear in the 
vicious walk model?  The GOE universality 
class is realized when walkers take $t$ 
steps under the nonintersecting condition. 
One possibility to observe the GUE class 
is to impose an additional condition that 
the nonintersecting walkers come back 
to the original position after $2t$ 
steps\cite{BAIK,NFV,BR2}. 
In analyzing the diffusion scaling limit of the 
vicious walk model, Katori and Tanemura\cite{KT1,KT2} 
recently noticed another possibility. 
The walker distribution depends 
not only on the observation time 
(the number of steps) $t$ but also 
on the time interval $T$ in which the 
nonintersecting condition is imposed. 
Suppose that all the walkers start from 
the origin at time $t=0$. While the ratio 
of the time $t$ and the nonintersection 
time interval $T$, $t/T$, is small, 
the walker distribution is asymptotically 
described by GUE. When the end of the 
nonintersection time interval arrives, 
namely at $t/T=1$, the distribution becomes 
identical to that of GOE. This means that, 
while walkers randomly move in the time interval 
between $t=0$ and $t=T$, a transition from GUE 
to GOE takes place.
\par
In this Letter we analyze the dynamical 
correlations among the vicious walkers 
in the transition region. Utilizing 
the equivalence to a multimatrix model 
in quantum field theory, we evaluate the 
dynamical correlation functions among 
arbitrary number of walkers 
at arbitrary number of times. The asymptotic 
limit of the large number of walkers 
will be evaluated and we will find that the 
transition becomes discontinuous.
\par
Let us consider $N$ independent symmetric simple random 
walks on ${\bf Z} = \{\cdots,-2,-1,0,1,2,\cdots\}$ 
started from $N$ distinct positions $2 s_1 < 2 s_2 
< \cdots < 2 s_N, \ s_j \in {\bf Z}$. The position of 
the $j$-th random walker at time $k \geq 0$ is denoted by 
$R^{s_j}_k$ and we impose the nonintersecting condition 
\begin{equation}
R^{s_1}_k < R^{s_2}_k < \cdots < R^{s_N}_k, \ \ \ 1 \leq \forall k \leq K.
\end{equation}
If a possible random walk satisfies the condition (1), then it is 
called a vicious walk. Let $V(R^{s_j}_K=e_j)$ be the realization 
probability of the vicious walks, in which the $N$ walkers arrive 
at the positions $2 e_1 < 2 e_2 < \cdots < 2 e_N, \ e_j 
\in {\bf Z}$, at time $K$. 
\par
A simplification is obtained in the diffusion scaling limit: we set 
$K = Lt, s_j = \sqrt{L}x_j/2, e_j = \sqrt{L}y_j/2$ 
and take the limit $L \rightarrow \infty$. In Ref.\cite{KT1}, 
it is clarified that $\lim_{L \rightarrow \infty} 
(\sqrt{L}/2)^N V(R^{\sqrt{L}x_j/2}_{Lt} = 
\sqrt{L}y_j/2) = f(t;\{y_j\} \mid \{x_j\})$, where
\begin{equation}
f(t;\{y_j\} \mid \{x_j\}) = \frac{1}{(2 \pi t)^{N/2}}
\det[e^{-(x_j - y_k)^2/2t}]_{j,k=1,\cdots,N}.
\end{equation} 
This function shows how the nonintersecting probability 
of the Brownian particles on the rescaled lattice ${\bf Z}/(\sqrt{L}/2)$ 
up to time $t$ depends on the initial and final positions. Therefore, 
for the nonintersecting Brownian motions in the time interval 
$[0,T]$, the transition density from the configuration 
$\{x_j\}$ at time $s$ to $\{y_j\}$ at time $t$ is given by\cite{KT2} 
\begin{equation}
\varphi^{T}(s,\{x_j\};t,\{y_j\}) = \frac{f(t-s;\{y_j\} \mid \{x_j\}) 
{\cal N}(T-t;\{y_j\})}{{\cal N}(T-s;\{x_j\})},
\end{equation}
where ${\cal N}(t;\{x_j\}) = \int_{y_1<\cdots<y_N} \prod_{j=1}^N dy_j f(t;\{y_j\} \mid \{x_j\})$. 
Note that there is an obvious temporal inhomogeneity since RHS depends not only 
$t-s$ but also $T-s$ and $T-t$. 
\par
The dynamics of the Brownian particles is described by the dynamical 
correlation functions. Let us denote the positions of the Brownian 
particles at a time $t_j$ by $x^j_1,x^j_2,\cdots,x^j_N$. 
Then the dynamical correlation functions among the particles 
at times $t_1,t_2,\cdots,t_M$ are defined as 
\begin{eqnarray} 
& & \rho(x^1_1,\cdots,x^1_{n_1};\cdots;x^M_1,\cdots,x^M_{n_M}) 
\nonumber \\ & = & \int \prod_{j=1}^N dx^0_j \int \prod_{j=n_1 + 1}^N dx^1_j 
\cdots \int \prod_{j=n_M + 1}^N dx^M_j \nonumber \\ 
& \times & p_0(\{x^0_j\}) \prod_{m=0}^{M-1} \varphi^{T}(t_m,\{x^m_j\};t_{m+1}, 
\{x^{m+1}_j\}).
\end{eqnarray}
Here $p_0(\{x^0_j\})$ is the initial distribution at $t_0=0$. Let us now 
suppose that all the particles start at the origin. Namely, we set 
$p_0(\{x^0_j\}) = \prod_j \delta(x^0_j)$ and obtain
\begin{eqnarray} 
& & \rho(x^1_1,\cdots,x^1_{n_1};\cdots;x^M_1,\cdots,x^M_{n_M}) 
\nonumber \\ & \propto & \int \prod_{j=n_1 + 1}^N dx^1_j 
\cdots \int \prod_{j=n_M + 1}^N dx^M_j \int \prod_{j=1}^N dx^{M+1}_j 
\nonumber \\ & \times & \prod_{j>k}^N (x^1_j - x^1_k) 
\prod_{j>k}^N {\rm sgn}(x^{M+1}_j - x^{M+1}_k) 
\nonumber \\ & \times & 
\prod_{m=1}^M \det[g^m(x^m_j,x^{m+1}_k)]_{j,k=1,\cdots,N}.
\end{eqnarray}
Here ($t_{M+1} \equiv T$) 
\begin{eqnarray}
g^1(x,y) & = & \frac{e^{-x^2/(2 t_1)} e^{-(x-y)^2/(2(t_2-t_1))}}
{\sqrt{2 \pi t_1} \sqrt{2 \pi(t_2-t_1)}}, \nonumber \\ 
g^m(x,y) & = & \frac{e^{-(x-y)^2/(2(t_{m+1}-t_{m}))}}
{\sqrt{2 \pi (t_{m+1}-t_{m})}}, \ \ 2 \leq  m \leq M. 
\end{eqnarray}
\par
At this stage we notice a direct correspondence between our problem and 
the multimatrix models in quantum field theory. Itzykson and Zuber firstly 
analyzed a two matrix model in which two hermitian random matrices 
were combined\cite{IZ,MEHTA}. Mehta and Pandey then coupled a 
real-symmetric and a hermitian random matrices to devise a $(1+1)$ 
matrix model, as a mathematical interpolation of GOE and GUE\cite{MP,PM}. 
After it had been realized that multimatrix models were useful in the quantum 
field theory on random surfaces\cite{DGZJ}, Eynard and Mehta invented a 
method to evaluate the correlation functions among the eigenvalues of 
combined $M$ hermitian matrices\cite{EM}. As a further generalization, 
Nagao proposed an $(M+1)$ matrix model in which one real symmetric and 
$M$ hermitian matrices were combined and showed that the 
correlation functions were generally written in the forms of quaternion 
determinants\cite{NAGAO}. We can readily see that the above dynamical 
correlation functions among vicious walkers have the same forms as 
the eigenvalue correlation functions of the $(M+1)$ matrix model.
\par
For simplicity we set $N$ even and summarize the quaternion determinant 
formulas in the following. In terms of ($x_m \equiv x,x_n \equiv y$) 
\begin{eqnarray}
& & G^{mn}(x,y) \nonumber \\ 
& = & \left\{ \begin{array}{l} \delta(x-y), \ \ m=n, \\ 
g^m(x,y), \ \ m=n-1, \\  
\displaystyle \int \prod_{j=m+1}^{n-1} dx_j 
\prod_{l=m}^{n-1}
g^l(x_l,x_{l+1}), \ \ m < n - 1, \end{array} \right.  
\end{eqnarray}
we firstly define 
\begin{eqnarray}
& & F^{mn}(x,y) \nonumber \\ & = & 
\int_{-\infty}^{\infty} dz^{\prime} 
\int_{-\infty}^{z^{\prime}} dz 
\{ G^{m \ M+1}(x,z) G^{n \ M+1}(y,z^{\prime}) \nonumber \\ 
& - & G^{n \ M+1}(y,z) G^{m \ M+1}(x,z^{\prime}) \}.
\end{eqnarray}
Let us introduce an antisymmetric inner product
\begin{eqnarray}
& & \langle f(x), g(y) \rangle \nonumber \\ 
& = & \frac{1}{2} \int dx \int dy  
F^{11}(y,x)  [f(y) g(x) - f(x) g(y) ]
\end{eqnarray}
and construct monic polynomials $R^1_k(x)=x^k+\cdots$ of 
degrees $k$ so that they satisfy the skew 
orthogonality relations:
\begin{eqnarray} 
& & \langle R^1_{2j}(x), R^1_{2l+1}(y) \rangle = 
- \langle R^1_{2l+1}(x), R^1_{2j}(y) \rangle = 
r_j \delta_{jl}, \nonumber \\ 
& & \langle R^1_{2j}(x), R^1_{2l}(y) \rangle 
= 0, \ \langle R^1_{2j+1}(x), R^1_{2l+1}(y) \rangle = 0.
\end{eqnarray}
We then define functions $R^m_k(x)$ and 
$\Phi^m_k(x)$, $m=2,3,\cdots,M+1$, as
\begin{eqnarray}
R^m_k(x) & = & \int dy R^1_k(y) G^{1 m}(y,x), \nonumber \\ 
\Phi^m_k(x) & = & \int dy R^m_k(y) F^{m m}(y,x).
\end{eqnarray}
Now matrices $D^{mn}$, $I^{mn}$ and $S^{mn}$ 
are introduced as
\begin{displaymath} 
D^{mn}_{jl} =  \sum_{k=0}^{(N/2)-1} \frac{R^m_{2 k}(x^m_j) R^n_{2 k + 1}(x^n_l) - R^m_{2 k + 1}(x^m_j) 
R^n_{2 k}(x^n_l)}{r_k}, 
\end{displaymath}
\begin{displaymath} 
I^{mn}_{jl} = \sum_{k=0}^{(N/2)-1} \frac{
\Phi^m_{2 k + 1}(x^m_j) \Phi^n_{2 k}(x^n_l)
- \Phi^m_{2 k}(x^m_j) \Phi^n_{2 k + 1}(x^n_l)}{r_k}, 
\end{displaymath}
and
\begin{displaymath}
S^{mn}_{jl} = \sum_{k=0}^{(N/2)-1} 
\frac{\Phi^m_{2 k}(x^m_j) 
R^n_{2 k + 1}(x^n_l) -  \Phi^m_{2 k + 1}(x^m_j) R^n_{2 k}(x^n_l)}{r_k}.  
\end{displaymath}
Further we define
\begin{displaymath}
{\tilde S}^{mn}_{jl} = \left\{ \begin{array}{l} S^{mn}_{jl}, \ \ m \geq n, \\ 
S^{mn}_{jl} - G^{mn}(x^m_j,x^n_l), \ \ m < n \end{array} \right. 
\end{displaymath}
and
\begin{displaymath}
{\tilde I}^{mn}_{jl} = I^{mn}_{jl} + F^{mn}(x^m_j,x^n_l).
\end{displaymath}
\par
For a self dual $N \times N$ quaternion matrix $Q = [q_{jk}]$, 
a determinant ${\rm Tdet} Q$, originally introduced into 
random matrix theory by Dyson\cite{DYQ}, is defined as 
\begin{equation} 
{\rm Tdet}\, Q = \sum_P (-1)^{N-l} \prod_1^l {\rm tr}(q_{ab} 
q_{bc} \cdots q_{da}), 
\end{equation}
where $P$ denotes any permutation of the indices 
$(1,2,\cdots,N)$ consisting of $l$ exclusive cycles of the 
form $(a \rightarrow b \rightarrow c \rightarrow \cdots \rightarrow 
d \rightarrow a)$ and $(-1)^{N-l}$ is the parity of $P$. Note 
that ${\rm tr}q$ is equal to a half of the trace of the $2 \times 2$ 
matrix representation of $q$. 
\par
Let us suppose that the elements of quaternion matrices $B^{\mu \nu}$, 
$\mu,\nu=1,2,\cdots,M$, have the following $2 \times 2$ 
representations:
\begin{equation}
B^{\mu \nu}_{jl} =  \left[ \begin{array}{cc} {\tilde S}^{\mu \nu}_{jl} & 
{\tilde I}^{\mu \nu}_{jl} \\ 
D^{\mu \nu}_{jl} & {\tilde S}^{\nu \mu}_{lj} \end{array} \right], 
\ \ j,l = 1,2,\cdots,N.
\end{equation}
Then Nagao's result\cite{NAGAO} asserts that
\begin{eqnarray}
& & \rho(x^1_1,\cdots,x^1_{n_1};\cdots;x^M_1,
\cdots,x^M_{n_M}) \nonumber \\ & \propto & 
{\rm Tdet}[B^{\mu \nu}(n_{\mu},n_{\nu})], \ \ \mu,\nu=1,2,\cdots,M,
\label{QDET}
\end{eqnarray}
where each block $B^{\mu \nu}(n_{\mu}, n_{\nu}) $ is obtained by removing 
the $n_{\mu} + 1, n_{\mu} + 2,  \cdots, N$-th rows and $n_{\nu} + 1, n_{\nu} 
+ 2, \cdots, N$-th columns from $B^{\mu \nu}$.  
\par
Let us examine the consequences of the quaternion determinant formula. 
Now we remark that the skew orthogonal polynomials $R^1_k(x)$ are 
explicitly written as
\begin{equation}
R^1_k(x) = \xi^{k/2} 
\sum_{j=0}^k \alpha_{kj} H_j \left( \frac{x}{c_1} \right) 
\xi^{-j/2}, 
\end{equation}
where $\xi = t_1/(2 T - t_1)$,
\begin{displaymath}
c_n = \sqrt{t_n (2 T - t_n)/T},
\end{displaymath}
\begin{eqnarray}
\alpha_{2 k \ j} & = & 2^{- 2 k} c_1^{2 k} \delta_{2 k \ j}, \nonumber \\ 
\alpha_{2 k + 1 \ j} & = &  2^{- 2 k - 1} c_1^{2 k + 1}( 
\delta_{2 k + 1 \ j} - 4 k \delta_{2 k - 1 \ j}) \nonumber 
\end{eqnarray}
and $H_j(x)$ are the Hermite polynomials. 
\par
We begin with the simplest case $M=1$ and $n_1=1$. Putting the explicit 
formula of $R^1_k(x)$ into the quaternion determinant expression and utilizing 
the asymptotic formula for the Hermite polynomials, we can readily obtain
\begin{equation}
\rho(x^1_1) \propto \frac{1}{\pi c_1} \sqrt{2 N - (x^1_1/c_1)^2}, \ \ 
\mid x^1_1 \mid < \sqrt{2 N} c_1
\end{equation}
in the limit $N \rightarrow \infty$. Thus the walker density always 
has a semicircle shape (Wigner's semicircle 
law), while the width of the semicircle is dependent on time and scaled by 
$c_1$. This result suggests that, after introducing a new rescaled 
variables $\lambda^m_j = x^m_j/c_m$, the vicious walk model in the 
diffusion scaling limit becomes equivalent to the matrix Brownian 
motion model\cite{DYB,FN,NFT,NFL,NFQ} which is normalized so that 
the width of the semicircle is a constant.
\par
The asymptotic forms of the dynamical correlation functions of 
matrix Brownian motion models were evaluated by Forrester, Nagao and 
Honner\cite{FNH}. In particular, in the edge region of the 
semicircle, dynamical asymptotic correlations are described by 
the Airy function ${\rm Ai}(x)$. We can directly reinterpret 
Forrester, Nagao and Honner's result in the context of vicious 
walk model. Let us introduce rescaled temporal 
and spatial variables $\tau_m$ and $X^m_j$ as 
\begin{eqnarray}
t_m & = &  \left( 1 - \frac{\tau_m}{N^{1/3}} \right) T, \nonumber \\ 
x^m_j & = & c_m \left(\sqrt{2 N} + \frac{X^m_j}{2^{1/2} N^{1/6}} \right)
\label{SCALING}
\end{eqnarray}
and take the limit $N \rightarrow \infty$. With an appropriate 
normalization, the dynamical correlation functions asymptotically 
have the same forms as in eq. (\ref{QDET}) except that the quaternion 
matrices $B^{\mu \nu}_{jl}$ are replaced by 
\begin{equation}
{\cal B}^{\mu \nu}_{jl} =  \left[ \begin{array}{cc} {\tilde 
{\cal S}}^{\mu \nu}_{jl} & 
{\tilde {\cal I}}^{\mu \nu}_{jl} \\ 
{\cal D}^{\mu \nu}_{jl} & {\tilde {\cal S}}^{\nu \mu}_{lj} \end{array} \right], 
\ \ j,l = 1,2,\cdots,N,
\end{equation}
where
\begin{eqnarray}
{\cal D}^{\mu \nu}_{jl} & = & \frac{1}{4} \left[ 
\int_0^{\infty} ds e^{-\tau_{\mu} s} {\rm Ai}(X^{\mu}_j + s) 
\frac{d}{ds} \{ e^{-\tau_{\nu} s} 
{\rm Ai}(X^{\nu}_l + s) \} 
\right. \nonumber \\ 
& - & \left. 
\int_0^{\infty} ds e^{-\tau_{\nu} s} {\rm Ai}(X^{\nu}_l + s) 
\frac{d}{ds} \{ e^{-\tau_{\mu} s} 
{\rm Ai}(X^{\mu}_j + s) \} 
\right], \nonumber 
\end{eqnarray} 
\begin{eqnarray}
{\tilde {\cal I}}^{\mu \nu}_{jl} & = & 
\int_0^{\infty} ds e^{-\tau_{\nu} s} {\rm Ai}(X^{\nu}_l - s) 
\int_s^{\infty} dv e^{-\tau_{\mu} v} {\rm Ai}(X^{\mu}_j - v) 
\nonumber \\ & - &  
\int_0^{\infty} ds e^{-\tau_{\mu} s} {\rm Ai}(X^{\mu}_j - s) 
\int_s^{\infty} dv e^{-\tau_{\nu} v} {\rm Ai}(X^{\nu}_l - v) 
\nonumber 
\end{eqnarray} 
and
\begin{displaymath}
{\tilde {\cal S}}^{\mu \nu}_{jl} = \left\{ 
\begin{array}{l} 
{\cal S}^{\mu \nu}_{jl}, \ \ \mu \geq \nu, \\  
{\cal S}^{\mu \nu}_{jl} - {\cal G}^{\mu \nu}_{jl}, \ \ \mu < \nu
\end{array} \right. 
\end{displaymath}
with
\begin{eqnarray}
{\cal S}^{\mu \nu}_{jl} & = &   
\int_0^{\infty} ds e^{(\tau_{\mu}-\tau_{\nu}) s} 
{\rm Ai}(X^{\mu}_j + s) 
{\rm Ai}(X^{\nu}_l + s) \nonumber \\ 
& + & \frac{1}{2} {\rm Ai}(X^{\nu}_l) \int_0^{\infty} ds 
e^{-\tau_{\mu} s} {\rm Ai}(X^{\mu}_j - s),
\nonumber 
\end{eqnarray} 
\begin{displaymath}
{\cal G}^{\mu \nu}_{jl} =  
\int_{-\infty}^{\infty} ds e^{(\tau_{\mu}-\tau_{\nu}) s} 
{\rm Ai}(X^{\mu}_j + s) 
{\rm Ai}(X^{\nu}_l + s).
\end{displaymath} 
\par
From the above asymptotic result we can extract information on 
limiting behavior. Let us firstly take the limit 
$\tau_{\mu} \rightarrow \infty$ with the time differences 
$\tau_{\mu}-\tau_{\nu}$ fixed. It can be readily seen that 
this limiting procedure is equivalent to set 
${\cal D}^{\mu \nu}_{jl}$, ${\tilde {\cal I}}^{\mu \nu}_{jl}$ 
and the second term of ${\cal S}^{\mu \nu}_{jl}$ zeros. 
Then the quaternion determinant is reduced to an 
ordinary determinant and the asymptotic correlation 
functions become temporally homogeneous. 
We find that they are the dynamical correlation functions 
within the universality class of GUE. The equal time 
correlation functions are described by the Airy 
kernel\cite{PJF,TW} in the temporally homogeneous region. 
Because of the rescaling (\ref{SCALING}), it can be seen that the GUE 
universality class survives until time $t$ very close 
to $T$: only when $T-t \sim O(N^{-1/3})$, the transition 
to GOE class occurs. Therefore we can conclude that 
the transition from GUE to GOE class is discontinuous 
in the limit $N \rightarrow \infty$.
\par
The second interesting case is $X^{\mu}_j \rightarrow - \infty$ 
with the position differences $X^{\mu}_j - X^{\nu}_l$ fixed. 
This reproduces the asymptotic dynamical correlation functions 
in the bulk region. They are spatially homogeneous and 
the equal time correlations are equivalent to 
Pandey and Mehta's asymptotic result\cite{PM}. 
In this bulk region we can also observe the discontinuous 
transition from GUE to GOE. 
\par
In summary, nonintersecting Brownian motion of $N$ 
particles in finite time interval $0 \leq t \leq T$ was 
studied in one dimension. Dynamical correlation functions 
among many particles at many times were written in the 
forms of quaternion determinants. The asymptotic forms of 
the dynamical correlation functions in the limit 
$N \rightarrow \infty$ were evaluated and the 
universality class transition from GUE to GOE was 
found to be discontinuous. The compact asymptotic formula 
was derived above only after taking the diffusion 
scaling limit in which much information contained 
in the original lattice model was suppressed. 
In order to fully understand the lattice vicious walk model, 
we need to study other scaling limits as well. Further 
studies in this direction should be made in future works.

\widetext
\end{document}